\documentclass{llncs}
\usepackage[utf8]{inputenc}
\usepackage{hyperref}
\usepackage[hmarginratio=1:1,top=25mm,bottom=25mm,left=20mm,right=20mm,columnsep=30pt]{geometry} 
\usepackage{graphicx}
\usepackage{float}
\usepackage{verbatim}
\usepackage{subcaption}

\usepackage[style=nature]{biblatex}

\title{Inspiration, Captivation, and Misdirection: Emergent Properties in Networks of Online Navigation}
\institute{Oxford Internet Institute, University of Oxford, 1 St Giles, Oxford, UK, OX1 3JS\and
Alan Turing Institute, The British Library, 96 Euston Road, London, UK, NW1 2DB}

\author{Patrick Gildersleve\inst{1} \& Taha Yasseri\inst{1, 2}}
\date{July 2017}

\begin{document}
\maketitle





\begin{abstract} 
The World Wide Web (WWW) has fundamentally changed the ways billions of people are able to access information. Thus, understanding how people seek information online is an important issue of study. Wikipedia is a hugely important part of information provision on the web, with hundreds of millions of users browsing and contributing to its network of knowledge. The study of navigational behaviour on Wikipedia, due to the site's popularity and breadth of content, can reveal more general information seeking patterns that may be applied beyond Wikipedia and the Web. 
Our work addresses the relative shortcomings of existing literature in relating how information structure influences patterns of navigation online. We study aggregated clickstream data for articles on the English Wikipedia in the form of a weighted, directed navigational network. We introduce two parameters that describe how articles act to source and spread traffic through the network, based on their in/out strength and entropy. From these, we construct a navigational phase space where different article types occupy different, distinct regions, indicating how the structure of information online has differential effects on patterns of navigation. Finally, we go on to suggest applications for this analysis in identifying and correcting deficiencies in the Wikipedia page network that may also be adapted to more general information networks.
\end{abstract}

\section{Introduction}

The Internet and particularly the World Wide Web (WWW) has brought a vast sea of information to the fingertips of billions of people, fundamentally changing the ways that we seek and gain information. Given the scale and importance of WWW it is important that people are able to navigate through it effectively, thus understanding how people seek information online is vital for the design of such information systems. Affordances of the platform (including webpage content, webpage design, and website hyperlink structure) and user desires determine whether users are assisted, misdirected, or manipulated in navigation. We may record traces of how users navigate online in clickstream logs - the sequence of clicks a user makes within and between webpages. These navigational data are an important representation for the quality and utility of a website, and may be harnessed by webmasters for improvements in website design \cite{meiss2010agents,Wu2014}. 

Wikipedia, the free, online, collaborative encyclopedia has become a hugely important part of information provision on WWW. As the web's fifth most popular website, it is the largest and most popular general reference work on the internet \cite{alexa2017}. Whilst it is not immune to criticism on its accuracy \cite{waters2007you}, biases \cite{wagner2016women}, and coverage \cite{samoilenko2014distorted} it has been meaningfully compared to the Encyclopedia Britannica \cite{giles2005}. Analysing clickstream data for Wikipedia is of particular interest, since we may study navigational behaviour for its hundreds of millions of users across the huge collaboratively generated network of knowledge. These data on navigation between the vast number of articles can reveal general patterns of information seeking behaviour as well as the influence of the article network's structure. These insights can be used by both the Wikimedia Foundation and editors to improve the website such as in regular error correction, more fundamental website design changes as well as in addressing important issues such as how regular users are affected by imbalance and systemic bias \cite{IJoC777, callahan2011, lam2011} of content across Wikipedia. More generally, this analysis can help us to understand human information seeking patterns beyond Wikipedia and even WWW.

Analysis of clickstream data, in the form of web usage mining, has been practised with the aim of improving website design and targeting users more effectively to increase use of the service on offer. For example; online shopping websites track users to understand the patterns of behaviour that may lead to a purchase \cite{Bucklin2002}, online social networks analyse navigation patterns within their website to provide customised experiences to increase user interaction and retention \cite{Benevenuto2009, Wang2016}, and clickstream data may be used to detect fake or automated accounts on online services \cite{Wang2016}.

Past work that covers more general patterns of navigation on WWW has tried to identify how people use the web (both within and between websites and individual pages), as compared to its design and structure. Weinreich et al. find that `Link following has remained the most common navigation action, accounting for about 45\% of all page transitions' \cite{Weinreich2006}. However, the structure of the web alone does not give a full picture of how it acts as a medium for the discovery of information for billions of people. Wu \& Ackland identify a `mismatch between hyperlinks and clickstreams' for navigation between the 1000 most popular websites (as ranked by Alexa.com), finding a marked difference between the network of hyperlinks and the navigational network of clickstreams \cite{Wu2014}. 
The authors comment that as we move through Web 2.0 \& Web 3.0 that this mismatch between hyperlinks and clickstreams will be alleviated, as users and algorithms ostensibly designed to serve users' interest (rather than individual webmasters) provide more relevant hyperlinks across the Web.

Wikipedia is one instance where users engagement in the form of editors' writing, correcting, and warring over articles shapes the form and structure of the website itself.
The editorial and traffic statistics of Wikipedia pages have been used to predict movies box office revenue \cite{mestyan2013early}, elections turnout and outcome \cite{yasseri2014can, yasseri2016wikipedia}, and disease outbreaks \cite{generous2014global}. 
Clickstream analysis can provide important insights that both the Wikimedia Foundation and editors may draw on when improving the website. For example, whilst many link prediction techniques rely solely on hyperlink structure and semantic features \cite{milne2008, noraset2014}, West et al. use navigation paths from the Wikispeedia game to create a link suggestion model \cite{West2015}. Unfortunately, this approach is limited to navigation where the eventual target page be explicitly known to both the algorithm and the user. A more general approach presented in \cite{Paranjape2016} utilises implicit signals from server logs to maximise objective functions under 3 web browsing models in suggesting the most useful links for the future, validated on the desktop English Wikipedia. 

Lamprecht et al. investigate how different naïve link selection models compare to data from the Wikipedia clickstream, finding that a model based on article structure best explains user navigation choices \cite{Lamprecht2016}. This work is built on by Dimitrov et al., by using a more complex model utilizing Bayesian inference, supplemented by mixed effects hurdle models using network, semantic and visual features to predict transition counts in clickstream data \cite{Dimitrov2017}. 
Finally, `Why We Read Wikipedia' \cite{Singer2017} provides a comprehensive overview of navigation on Wikipedia to create a taxonomy of users, their behaviours and their motivations by matching survey responses with data including user clickstreams. A wide range of navigational patterns are observed, including fast-paced random exploration, current events driven navigation, and long sessions of work and research.

The existing literature well covers the analysis of user behaviour from clickstream data including its applications in improving website design and user experience across the WWW and specific to Wikipedia. However, relatively little research covers the properties and utility of the navigational graph itself. As a complement to user focussed clickstream analysis, traces of user navigation from clickstream data may be used as part of a page-focused analysis. Put simply, instead of asking `how do users behave?', we shift the lens of focus to the webpages in order to ask `how are articles used?'. This change in reference frame allows us to directly investigate how the structure of information on Wikipedia (and the web at large) supports and hinders different kinds of navigational behaviour. Formally, our main contributions are to introduce two metrics to describe the sourcing and spreading of user traffic through webpages and to use these metrics to construct a phase space that is used to analyse patterns of user behaviour. Different page types introduce different, distinct navigational structures into the page network. We finish by recommending use cases for this navigational phase space in identifying errors and deficiencies in the Wikipedia page network.

\section{Results \& Discussion}

\begin{figure}
\centering
\begin{subfigure}{.2\textwidth}
\centering
\includegraphics[width=\linewidth]{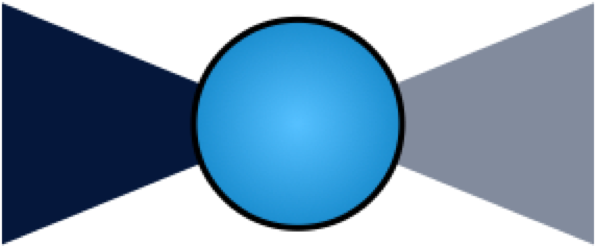}
\caption{Sink: $G<0$}
\label{fig:sink}
\end{subfigure}
\begin{subfigure}{.2\textwidth}
\centering
\includegraphics[width=\linewidth]{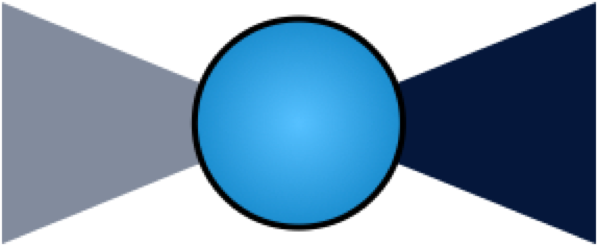}
\caption{Source: $G>0$}
\label{fig:source}
\end{subfigure}
\begin{subfigure}{.2\textwidth}
\centering
\includegraphics[width=\linewidth]{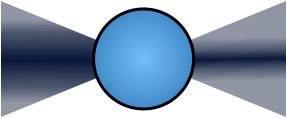}
\caption{Bottleneck: $D<0$}
\label{fig:bottleneck}
\end{subfigure}
\begin{subfigure}{.2\textwidth}
\centering
\includegraphics[width=\linewidth]{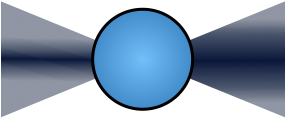}
\caption{Distributor: $D>0$}
\label{fig:distributor}
\end{subfigure}
\caption{An illustration of the different sourcing/spreading configurations. The intensity of shading schematically codes the volume of traffic across links to/from an article's neighbours.}
\label{fig:sourcespread}
\end{figure}

The Wikipedia clickstream dataset contains monthly aggregates for the number of clicks on hyperlinks between articles on Wikipedia. From a network perspective, these data act as an edge list and may be used to construct a weighted, directed graph. We use measures of how much a page acts as a relative traffic source or sink (sourcing coefficient $G$) and how much a page acts to spread or focus traffic (spreading coefficient $D$) as part of a `Navigational Phase Space' to describe and visualise how users navigate through pages on Wikipedia. Illustrations of different sourcing and spreading configurations are provided in Fig. \ref{fig:sourcespread}.

The sourcing coefficient of a given node (article) is defined as
\begin{equation}
    G = \frac{S_{\mathrm{out}}-S_{\mathrm{in}}}{S_{\mathrm{out}}+S_{\mathrm{in}}} \enspace,
\end{equation}
where $S_{\mathrm{in/out}}$ is the in/out strength of the node i.e. the total number of clicks into/out from the article. Flow is not conserved in this network, since users may arrive at a node from an external source and may also click on multiple links within one page, so $S_{\mathrm{out}}$ may be larger than $S_{\mathrm{in}}$ for a given node. 

\begin{figure}
\centering
\begin{subfigure}{0.32\textwidth}
\includegraphics[width=\linewidth]{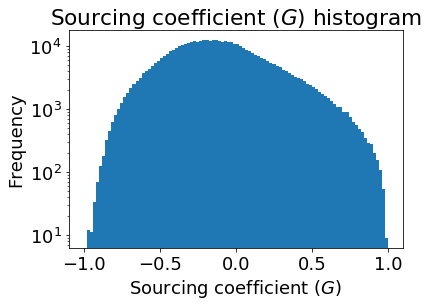}
\caption{}
\label{fig:sourcehist}
\end{subfigure}
\begin{subfigure}{0.32\textwidth}
\includegraphics[width=\linewidth]{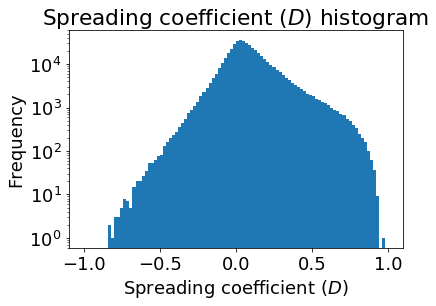}
\caption{}
\label{fig:spreadhist}
\end{subfigure}
\begin{subfigure}{0.32\textwidth}
\centering
\includegraphics[width=\linewidth]{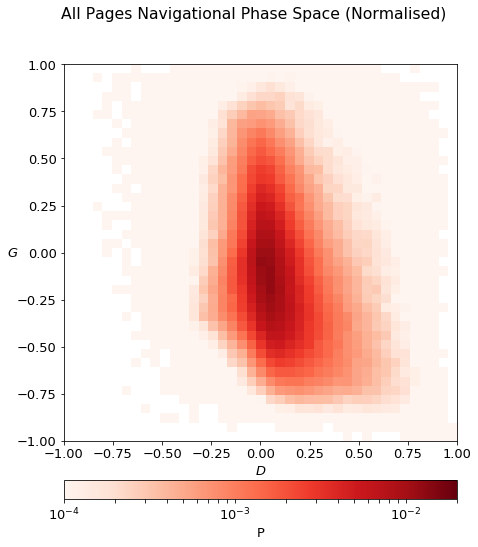}
\caption{}
\label{fig:phase}
\end{subfigure}
\caption{The distributions of all articles' sourcing coefficient (a) and spreading coefficient (b) are combined to form a navigational phase space (c). $P$ is the probability of an article existing at that point in phase space}
\end{figure}

The distribution of $G$ for all articles in the month of September 2016 is shown in Fig. \ref{fig:sourcehist}. A wide range of values is observed for $G$. However, on the whole, more pages act as traffic sinks ($G<0$) rather than sources ($G>0$) since users eventually stop browsing.

How the shape of traffic changes as it passes through nodes is another important emergent feature for each article. An article may receive disperse traffic from a range of neighbours or focussed traffic from a narrow subset of neighbours and then go on to send disperse or focussed traffic. 

The spreading coefficient is defined as
\begin{equation}
    D = \bar \sigma_{\mathrm{out}} - \bar \sigma_{\mathrm{in}},
\end{equation}
where $\bar \sigma_{\mathrm{in / out}}$ is the normalised in/out entropy of the node. This describes the spread of traffic, whether it is dispersed across many neighbours, or focussed over a relatively narrow subset of neighbours, for further details on node entropy see Sec.~\ref{sec:datamethods}.

The distribution of $D$ for articles in September 2016 is shown in Fig. \ref{fig:spreadhist}. We note that most articles act as `channels' ($D\sim0$), with no noticeable effect on the shape of traffic and that more articles act as distributors ($D>0$), than bottlenecks ($D<0$). 

We use these measures to construct a 2D navigational phase space for articles, shown in Fig. \ref{fig:phase}. The point that an article exists in within this space describes the nature of the traffic through it.

\begin{figure}
\begin{subfigure}{.5\textwidth}
\centering
\includegraphics[width=\linewidth]{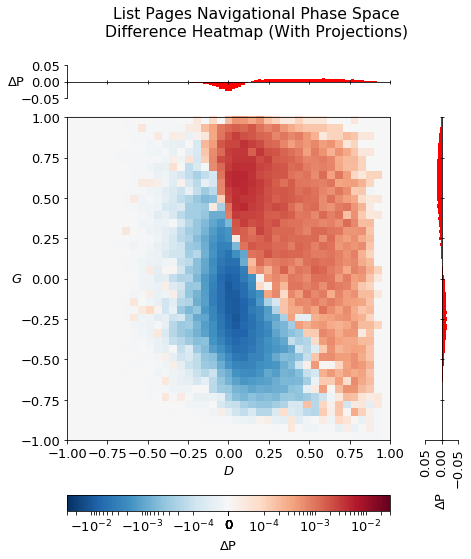}
\caption{}
\label{fig:list}
\end{subfigure}
\begin{subfigure}{.5\textwidth}
\centering
\includegraphics[width=\linewidth]{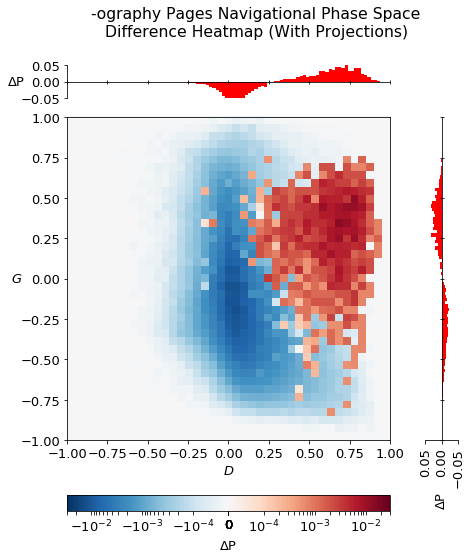}
\caption{}
\label{fig:ography}
\end{subfigure}

\begin{subfigure}{.5\textwidth}
\centering
\includegraphics[width=\linewidth]{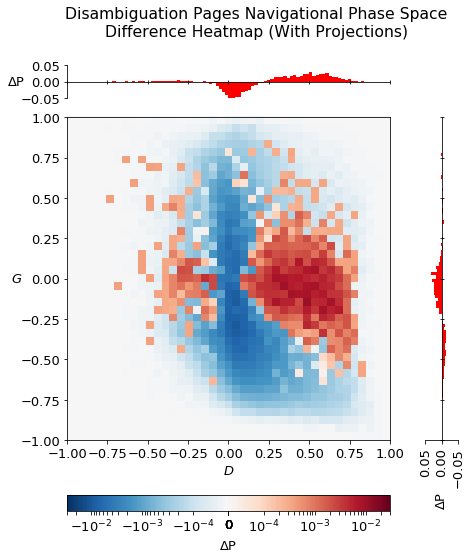}
\caption{}
\label{fig:disambiguation}
\end{subfigure}
\begin{subfigure}{.5\textwidth}
\centering
\includegraphics[width=\linewidth]{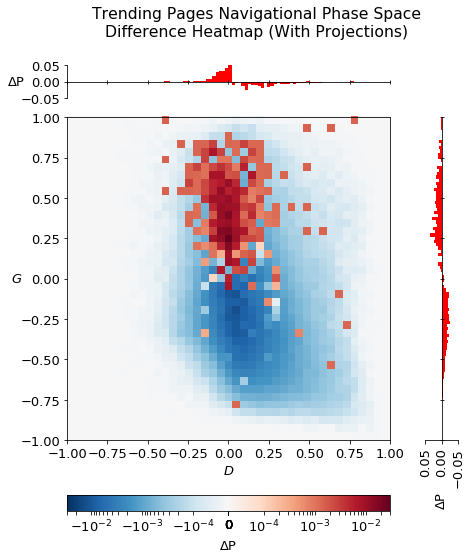}
\caption{}
\label{fig:trending}
\end{subfigure}
\caption{Navigational Phase Space Difference Heatmaps for a) List Pages, b) -ography pages, c) Disambiguation pages, d) Trending pages. $\Delta P$ is the difference between the values of the probability distributions at that point for a given page type and the distribution for all articles. Distinct patterns of navigation are observed for different article types.}\label{fig:phase_all}
\end{figure}

Considering the sourcing and spreading coefficients for each article on the English Wikipedia, we build the navigational phase spaces for 4 selected different page types - list pages, -ography pages, disambiguation pages, and trending pages. List articles provide  a list of links to all articles of a particular class (e.g. \emph{List of common misconceptions}). -ography articles act to summarise the body of work of creative professionals (e.g \emph{Paul McCartney Discography, Leonardo DiCaprio Filmography}). Disambiguation articles exist to resolve any ambiguity when articles on several topics might be expected to have the same `natural’ page title (e.g. \emph{Wooden (disambiguation)}). Trending articles are the most popular articles that receive a peak in popularity over the month of study (for more details on article types see Sec. \ref{sec:pagetype}. By examining the difference between the navigational phase spaces for these particular article types against that for all Wikipedia articles we are able to identify distinct navigational patterns. The results are shown in Fig. \ref{fig:phase_all}.


List pages act as both relatively strong sources and distributors of traffic, as indicated in Fig. \ref{fig:list}. Users may arrive at a list page from an external website or from a relatively narrow range of Wikipedia articles and be inspired to open a wide range of articles from the list.

-ography pages (Fig. \ref{fig:ography}) act as strong sources and distributors of traffic, more so than regular list pages. Traffic towards –ography pages is predominantly focussed from the respective actor/author/band etc. and, as with list pages, users are inspired to open a wide range of articles from the –ography page. 

One would expect disambiguation pages to spread traffic to a range of articles, since they are designed to act as navigation aids when a user may be searching for any of a range of similarly titled articles. We indeed observe this with the mode of articles with $D>0$ in Fig. \ref{fig:disambiguation}. However, there is a second mode with $D<0$; disambiguation pages which act to focus traffic. This is atypical and likely undesirable behaviour, evidence of misdirection that is later elaborated on.

Trending articles (Fig. \ref{fig:trending}) and related / linked articles are popular, by definition, receiving a large amount of traffic, predominantly from external sources. It is of no surprise that these articles then act as strong sources of traffic to the rest of the article network, however, they do not act to focus or distribute traffic; captivated users come from and navigate towards a similar spread of articles.

Clearly, the way that information is structured on Wikipedia has differential effects on patterns of navigation on the website. Future research on the production and structure of information on Wikipedia must clearly detail the impact on regular users' navigational behaviour and ability to access information.

This navigational phase space may also be used to identify errors and deficiencies in the Wikipedia page network and in principle the network of pages of any website. Broadly, these are characterised by any of; `extreme' navigational behaviour, `atypical' behaviour, and `mimetic' behaviour.

\subsubsection{Extreme behaviour}

An article occupying an extreme position in the navigational phase space is usually indicative of some error relating to hyperlinks on the page or hyperlinks leading to the page. Sourcing coefficient $G\sim-1$ may be a result of an article being erroneously linked to by another, far more popular, page that distorts the data for traffic into the article. Spreading coefficient $D\sim-1$ would indicate that the vast majority of users seek the information of another article following or instead of the information on the current article. This information could be used to inform the merger of articles, or identify erroneous hyperlinks to the first page that should be targeted at the second.
An example is the article `Passenger' (\emph{a person who travels in a vehicle but bears little or no responsibility...}) - $D = -0.79$. A brief investigation reveals that the article is linked to by a template\footnote{a mini-page that can be automatically copied and updated across many articles} for `Nettwerk', a popular music label that manages `Passenger (singer)'. The result of this is that many articles on acts relating to the Nettwerk music label erroneously link to `Passenger' instead of `Passenger (singer)' and that users browsing acts relating to Nettwerk traverse this link to only end up at the page relating to people travelling in vehicles. At this point, after receiving incorrect information, most users' strategy is to click through to `Passenger (disambiguation)' (resulting in strong article bottleneck behaviour) which does link to `Passenger (singer)' - the desired page.

\subsubsection{Atypical behaviour}

As previously observed, certain page types typically occupy distinct areas in the navigational phase space. Aside from more general extreme behaviour, some articles of particular types may exhibit behaviour that is atypical and perhaps undesirable for their type.

Consider the aforementioned 2 mode distribution for disambiguation pages in Fig. \ref{fig:disambiguation}. Articles in the mode with $D<0$ act as bottlenecks, focussing traffic towards one page - suggesting that on the whole, users find no ambiguity and mostly require one page in particular. An example of this is for the page `Tinder (disambiguation)' ($D=-0.70$) from which traffic is heavily focussed towards `Tinder (app)', rather than the combustible material. This behaviour would suggest that a disambiguation page might not be necessary and that `Tinder (app)' should be the default.

\subsubsection{Mimetic behaviour}

Occasionally, a general article will behave very similarly to articles of a particular type without explicitly being named or set out as such. We have observed this most clearly with articles behaving like list pages. Examples of these include `Saturday Night Live cast members’ ($G =0.61 $, $D = 0.71$), `Allied leaders of World War II’ ($G =0.16 $, $D = 0.76$), and `Bollywood horror films’ ($G =0.96 $, $D = 0.41$). This could be the basis for simple name changes or even splitting these kind of articles into separate dedicated list and descriptive pages.


\section{Conclusion}

This work emphasises that information structure is an important factor for navigation online and that hyperlinks are not created equal. Moreover, how information is organised between and within pages shapes both the volume and shape of traffic that flows across hyperlinks between them. Structure in the information network translates non-trivially to patterns of user navigational behaviour so it is important that this information structure acts to appropriately direct users, rather than to manipulate or misdirect them. Analysing pages' sourcing and spreading behaviour as part of a navigational phase space can act to further inform approaches to incorporating insights from data on user desires into the production and design of content online.

We have provided several practical suggestions for using this analysis to improve the content and structure of Wikipedia. There is certainly further work that may utilise these methods for studying patterns of navigation on and improving Wikipedia and other websites. Similar analysis could be applied to how patterns of navigation vary amongst different categories of content (politics, religion, sport etc.), as well as deeper analysis on how navigation behaviour is related to properties of pages such as their popularity, rankings of article quality, and controversiality \cite{yasseri2012}. We hope that this work inspires further research and application of clickstream data to issues of information structure and navigation online, emphasising that work on web structure should not be conducted in isolation from assessing its impact on regular users' search for information.

\section{Data \& Methods}
\label{sec:datamethods}
\subsubsection{Data:}

\begin{figure}[H]
\centering
\includegraphics[width=0.4\linewidth]{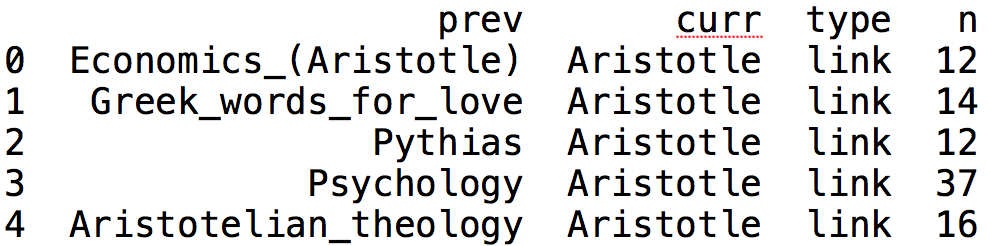}
\caption{A sample of Wikipedia clickstream data.}
\label{fig:csdata}
\end{figure}

The Wikipedia clickstream dataset \cite{clickstream} contains several month long counts of the total number of clicks between pairs of pages (referrer, resource), including traffic from external sites such as Google and Facebook, as well as the type of click (internal link, external link, redlink - for missing pages, other - for searches \& referrer spoofing). 
Measures to filter out activity from bots are enacted, and any (referrer, resource) pair with fewer than 11 clicks is cut from the dataset. An example of the records from one month are shown in Fig. \ref{fig:csdata}. From a network perspective, these data function as an edge list, detailing edges between source and target nodes with their respective weights. From this edge list we construct a weighted, directed navigation network of pages for 1 month of user navigation on Wikipedia. For the purpose of this project, we only consider navigation and information seeking behaviour across links within Wikipedia for the month of September 2016. This leaves us with a network that has 2,227,070 nodes and 13,951,247 edges. The total number of clicks (sum of all weights) in this dataset is 1,187,607,386.

%


\subsubsection{Node level statistics:}
\label{sec:nodestats}
For each article (node) in the network, we consider 3 basic properties for traffic both in and out; degree $k$, strength $S$, and entropy $\sigma$. In/out degree describes the number of neighbouring articles that an article receives traffic from/sends traffic to. In/out strength describes the total volume of traffic into / out of an article. Finally, a node's in/out entropy, given by
\begin{equation}
    \sigma_{\mathrm{in/out}} = -\sum_{i \in \mathrm{in/out \  edges}}\frac{w_i}{S_{\mathrm{in/out}}} \ln\left (\frac{w_i}{S_{\mathrm{in/out}}}\right ) \enspace,
\end{equation}
where $w_i$ is the weight of a given edge in/out, describes the spread of traffic into/out from a page. That is, whether traffic is focussed from/towards a narrow set of an article's neighbours or whether traffic from/towards an article is relatively evenly distributed across a wide set of its neighbours. This measure is normalised according to the maximum possible entropy for a page of given degree $k$;
\begin{equation}
    \bar \sigma = \frac{\sigma}{\ln(k)} \enspace.
\end{equation}

We note that the dataset filter for minimum edge weight (number of clicks) between articles introduces boundary effects for articles with low degree and low strength. For articles in the dataset with low degree and/or strength, any links they might have with slightly fewer than 11 clicks that are filtered out have a larger relative effect on that page's overall recorded properties. To mitigate these effects, whilst all edges in the dataset from internal links in a month long Wikipedia article network are present in our analysis, we do not study articles with in/out degree or strength below certain defined thresholds. Firstly, an article must have traffic both into and out from it. Secondly, we take the peak from the navigational network's in/out degree and strength distributions, and only consider articles with degree/strength above this value. Minimum values are typically $k \sim 2$ \& $S \sim 150$. The remaining set of articles (1376230 for September 2016) is where we focus our analysis. 

\subsubsection{Page type analysis:}  \label{sec:pagetype}
By analysing the strings of text of the page title we detect the article type: List pages (n=17997), Disambiguation pages (n=2379), -ography pages (n=1337). We also introduce a group of trending articles (n=651) based on the volume of external traffic towards an article - so as not to directly influence $S_{\mathrm{in}}$ which is based on internal traffic. A trending article must be one of the 5000 most popular articles in a particular month and must also receive its peak volume of traffic within said month compared to the other months in the dataset. This removes consistently popular articles, preserving those which receive a spike in popularity.

\begin{figure}[H]
\centering
\includegraphics[width=\linewidth]{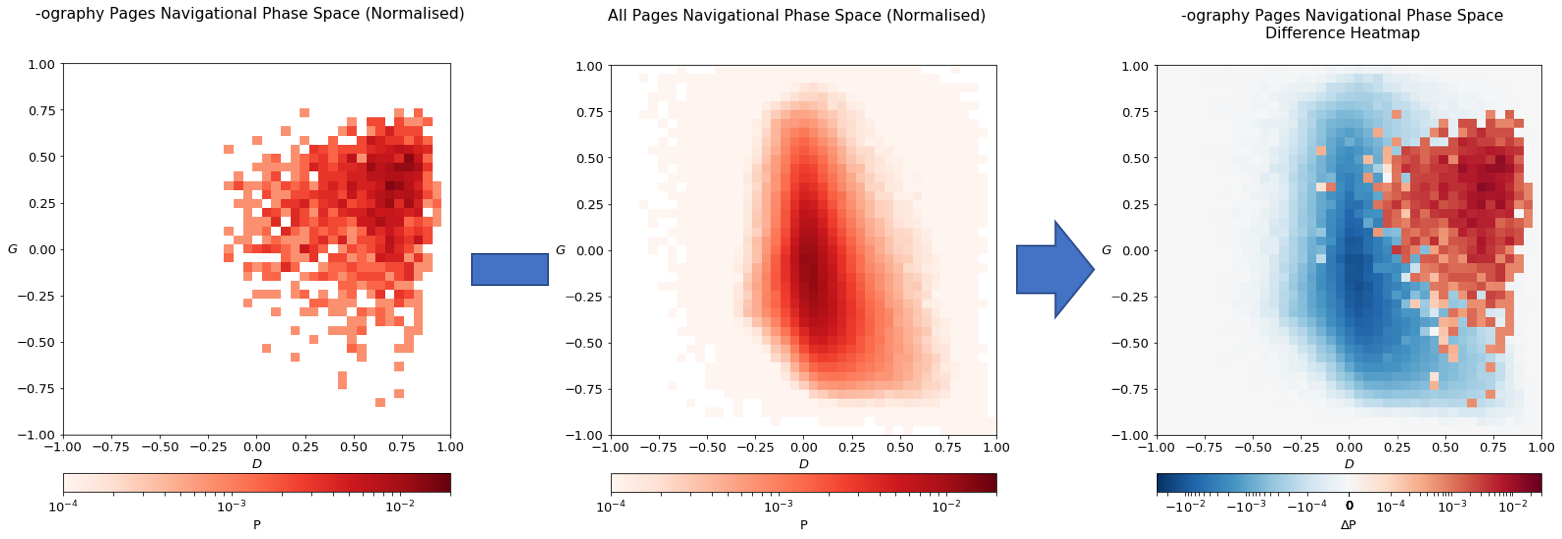}
\caption{By subtracting the navigational phase space for all articles from that for articles of a particular type, we highlight what patterns of navigational behaviour are more (or less) likely to be associated with articles of that particular type.}
\label{fig:process}
\end{figure}

For each of these article types, we find the distributions of sourcing parameter and spreading parameter, and form a normalised navigational phase space - equivalent to a 2D probability distribution. We then subtract the equivalent 2D probability distribution for all articles. This generates a heatmap highlighting where particular article types are more or less likely to exist as compared to the distribution of all Wikipedia articles. An example of this process is shown in Fig. \ref{fig:process}. This navigation phase space difference heatmap is our main tool for analysis of different types of Wikipedia article. 

\printbibliography
\end{document}